\documentclass[aps,pra,amsfonts,amssymb,amsmath,twocolumn,eqsecnum,showpacs,nofootinbib]{revtex4}

 \newtheorem{trial definition}{Trial definition}[section]

\begin{document}
 \title{The No Cloning Theorem versus the Second Law of Thermodynamics}
 \author{Gavriel Segre}
 \email{info@gavrielsegre.com}
 \homepage{http://www.gavrielsegre.com}
 \thanks{I would strongly to thank prof. Asher Peres for his intellectual honesty and his availability on fairly discussing the issue of this paper as well as for very useful suggestions}
 \bigskip
 \begin{abstract}
 Asher Peres' proof that a violation of the No Cloning Theorem would imply a
 violation of the Second Law of Thermodynamics is shown not to
 take into account the algorithmic-information's contribution to
 the Thermodynamical Entropy of the semi-permeable membranes of
 Peres' engine.
 \end{abstract}
 \maketitle
 \section{Introduction}
Two  results have changed, in the last two decades, our way of
looking at the  Foundation of Quantum Mechanics: the No-Cloning
Theorem  (by Dieks, Wooters and Zurek \cite{Peres-95},
\cite{Nielsen-Chuang-00}) and the comprehension of the
algorithmic-information's contribution to the thermodynamical
entropy in presence of Mawxell's demons (by Landauer, Bennett and
Zurek \cite{Landauer-90a}, \cite{Bennett-90a}, \cite{Bennett-90b},
\cite{Zurek-89}, \cite{Zurek-90a}, \cite{Zurek-90b}).

The No-cloning theorem, stating the impossibility of building a
quantum gate able to clone two non-orthogonal states, would seem
to have no connection with Quantum Thermodynamics; that this is
not the case, anyway, is implied by its equivalence with the
Theorem of Indistinguishability for nonorthogonal states lying at
the heart of the irreducibility of Quantum Information Theory to
the classical one.

In \cite{Peres-90} (as well as in the $ 9^{th} $ chapter of his
wonderful book \cite{Peres-95}) Asher Peres claims that the
Theorem of Indistinguishability for nonorthogonal states is
necessary in order of preserving the Second Law of
Thermodynamics; his proof of this statement is based on the
analysis of a cyclic thermodynamical engine  in which some
"magic" semi-permeable membranes, assumed ad absurdum to be able
to distinguish nonorthogonal states, are used in  a suitable way
in order of lowering the Universe's entropy.

As we will show, anyway, such a proof is not correct, since it
doesn't take into account the Landauer- Bennett- Zurek's results
on Mawxell's demon, that imply  that also the
algorithmic-information of Peres' semi-permeable membrane
contribute to the thermodynamical entropy, preventing the Second
Principle to be violated.

This consideration, already presented in the remark7.3.10 of my
PHD-thesis \cite{Segre-02}, is here extensively analyzed.

\section{No-cloning Theorem and indistinguishability of
nonorthogonal states} \label{sec:No-cloning Theorem and
indistinguishability of nonorthogonal states}

Let us us consider a quantum gate $ \hat{U} $ with two input edges
and two outputs edges such that there exist a normalized start
state $ | s
> $ and two distinct vectors $ | \psi_{1} > $ and  $ | \psi_{2} > $ such
that:
\begin{equation}\label{eq:cloning of the first state}
  \hat{U}  \,  | \psi_{1} > | s >   \; = \; | \psi_{1} > | \psi_{1} >
\end{equation}
and:
\begin{equation}\label{eq:cloning of the second state}
  \hat{U}  \,  | \psi_{2} > | s >   \; = \; | \psi_{2} > | \psi_{2} >
\end{equation}
Taking the inner product of eq.\ref{eq:cloning of the first
state} and eq.\ref{eq:cloning of the second state} one obtains
the equation:
\begin{equation}
  < \psi_{1} | \psi_{2} > ^{2} \; = \; < \psi_{1} | \psi_{2} >
\end{equation}
from which it follows that
\begin{equation}
  < \psi_{1} | \psi_{2} > \, = \, 0
\end{equation}
The No-Cloning Theorem, stating the impossibility of a device
able of cloning two nonorthogonal states, is then proved.

Let us now consider a different situation in which Alice codifies
her answer to Bob's marriage proposal sending him one of two
possible states $ | \psi_{1}
> $ and $ | \psi_{2} > $ (with the previously concorded rule that $ | \psi_{1}
> $ means \emph{yes} while $ | \psi_{2} > $ means \emph{no}). To
know  Alice's answer, Bob makes on the received state the
measurement described by the positive-operator-valued-measure
 $ \{ \hat{M} _{j} \}_{j=1}^{2} $ with outcome j. Depending on the
outcome on the measurement Bob tries to guess what the index i
was using some rule $ i \, = \, f(j) $, where $ f( \cdot ) $
represents the rule he uses to make the guess.

We will know prove the Theorem of Indistinguishability of
Nonorthogonal States stating that if $ | \psi_{1}
> $ and $ | \psi_{2} > $ are nonorthogonal it follows that Bob
cannot infer if Alice has accepted his marriage proposal.

Introduced the operators:
\begin{equation}
  \hat{E}_{i} \; := \; \sum_{j : f(j) = i} \hat{M} _{j}^{ \dag } \hat{M} _{j}
\end{equation}
the condition that Bob can infer Alice's answer may be formalized
by the constraint:
\begin{equation} \label{eq:distinguihability's constraint}
  <  \psi_{i} | \hat{E}_{i} | \psi_{i} > \; = \; 1 \; \; i \, = \,
  1,2
\end{equation}
Since $ \sum_{i} \hat{E}_{i} \; = \; {\mathbb{I}} $ it follows
that $ \sum_{i} <  \psi_{1} | \hat{E}_{i} | \psi_{1} > \; = \; 1
$; assuming the distinguishability condition of
eq.\ref{eq:distinguihability's constraint} it follows that $ <
\psi_{1} | \hat{E}_{2} | \psi_{1} > \; = \; 0 $ and thus:
\begin{equation}\label{eq:mixed annihilation}
  \sqrt{\hat{E}_{2}} | \psi_{1} > \; = \; 0
\end{equation}
Let us now suppose ad absurdum that $ < \psi_{1} | \psi_{2} > \,
\neq \, 0 $.

It follows that there exist a state $ | \psi_{\bot} > $ and two
complex numbers $ \alpha $ and $ \beta $ such that:
\begin{align}
   | \psi_{2} >  &  \; =  \; \alpha | \psi_{1} > \, + \, \beta | \psi_{\bot} > \\
  | \alpha |^{2} & \, + \, | \beta |^{2} \; = \; 1 \\
  | \beta | \; < \; 1 \\
 < \psi_{1} | \psi_{\bot} > \; = \; 0
\end{align}
Conseguentially:
\begin{equation}
   \sqrt{\hat{E}_{2}} | \psi_{2} > \; = \;  \beta
   \sqrt{\hat{E}_{2}} | \psi_{\bot} >
\end{equation}
and hence:
\begin{multline}
  <  \psi_{2} | \hat{E}_{2} | \psi_{2} > \; = \; | \beta |^{2} \, <  \psi_{\bot} | \hat{E}_{2} | \psi_{\bot}
  >  \\
  \leq \; | \beta |^{2} \, \sum_{i} <  \psi_{\bot} | \hat{E}_{i} |
  \psi_{\bot} > \; = \; | \beta |^{2} \; < \; 1
\end{multline}
which contradicts the absurdum hypothesis.

Beside their apparent diversity, the No-Cloning Theorem and the
Theorem of Indistinguishability of Nonorthogonal States may be
easily proved to be equivalent:

if Bob was able to distinguish the two nonorthogonal states $ |
\psi_{1} > \, , \, | \psi_{2} > $ Alice used to answer his
marriage proposal, he could clone them by making the measurement
distinguishing them and making at will multiple copies of the
state Alice had given him.

Contrary, if the non-orthogonal states $ | \psi_{1} > \, , \, |
\psi_{2} > $ were clonable, Bob could easily distinguish them by
cloning them reapetedly in order of obtaining the states $ |
\psi_{1} >^{ \bigotimes n } \, , \, | \psi_{2} >^{ \bigotimes n }
$ whose inner product tends to zero when $ n \rightarrow \infty $
and are, conseguentially, asymptotically distinguishable by
projective measurements.

\section{Bennett's Theorem on Maxwell's Demons in Classical Thermodynamics} \label{sec:Bennett's Theorem on Maxwell's Demons in Classical Thermodynamics}
Almost all the greatest physicists of the last two centuries has,
at some point, fought against one of the deepest problems of
Thermodynamics: Maxwell's demon.

Let us  introduce it with Maxwell's own words:
\begin{center}
  \textit{"One of the best extablished facts in thermodynamics is that it is impossible in a system enclosed in an envelope
which permits neither change of volume nor passage of heat, and
in which both the temperature and the pressure are everywhere the
same, to produce any inequality of temperature or of pressure
without the expenditure of work. This is the second law of
thermodynamics, and it is undoubtedly true as long as we can deal
with bodies only in mass, and have no power of perceiving or
handling the separate molecules of which they are made up. But if
we conceive a being whose faculties are so sharpened that he can
follow every molecule in its course, such a being, whose
attributes are still as essentially finite as our own, would be
able to do what is at present impossible to us. For we have seen
that the molecules in a vessel full of air at uniform temperature
are moving with velocities by no means uniform, though the mean
velocity of any great number of them, arbitrary selected, is
almost exactly uniform. Now let us suppose that such a vessel is
divided in two portions, A and B, by a division in which there is
a small hall, and that a being, who can see the individual
molecules, opens and closes this hole so as to allow only the
lower ones to pass from B to A. He will see, thus, without
expenditure of work, raise the temperature of B and lower that of
A, in contradiction with the second law of thermodynamics"; cited
from the last but one section "Limitation of the Second Law of
Thermodynamics" of the $22^{th}$ chapter of \cite{Maxwell-01} }
\end{center}
In the 220 years after the publication of Maxwell's book an
enormous literature tried to exorcize Maxwell's demon in different
ways; an historical review may be found in the first chapter
"Overview" as well as in the "Chronological Bibliography with
Annotations and Selected Quotations" of the wonderful books edited
by Harvey S. Leff and Andrew F. Rex  \cite{Leff-Rex-90} ,
\cite{Leff-Rex-03}.

All these exorcisms were based on the idea that, to accomplish
his task, Maxwell's demon necessarily causes a
thermodynamical-entropy's raising causing the Second Law to be
preserved:

they anyway strongly differed in identifying the element of the
demon's dynamical evolution which is \textbf{necessarily
thermodinamically-irreversible}:

coming to recent times, most of the  Scientific Community (not
only of Physics: cfr. e.g. the third chapter "Maxwell's Demons" of
\cite{Monod-72}) strongly believed in Leon Brillouin's exorcism
\cite{Brillouin-90}, identifying such an element in the
\textbf{demon's information-acquisition's process}.

When anyone thought that the "The-end" script had at last
appeared to conclude "The Exorcist" movie, Charles H. Bennett
showed in 1982 \cite{Bennett-90a}, \cite{Bennett-90b},
\cite{Feynman-96}, basing on the previous work by Rolf Landauer on
the Thermodynamics of Computation \cite{Landauer-90a}, that:
\begin{enumerate}
  \item Maxwell's Demon was still alive  owing to the
nullity of Brillouin's exorcism: the demon's acquisition process
may be done in a completelly thermodynamically-reversible way
  \item the necessarily-thermodinamically-irreversible element is instead demon's information-erasure's process
\end{enumerate}

The corner-stone of the Themodynamics of Computation is
Landauer's Principle:

in this framework an arbitrary function is called
logically-reversible if it is injective while it is called
thermodynamically-reversible if there exist a physical device
computing it in a thermodynamically-reversible way; Landauer's
Principle states the equivalence of logical-reversibility and
thermodynamical-reversibility.

An immediate consequence of Landauer's Principle is that the
erasure of information is thermodynamically-irreversible:

to prove it, it is sufficient to observe that to  any
logically-irreversible function one may associate a
logically-reversible function different from the original one in
that the output is augmented by some of the input's information
(usually called garbage); assuming ad absurdum that garbage's
erasure is thermodynamically-reversible, it would  then follow
that the original function would be thermodynamically-reversible
too, contradicting the hypothesis.

We can at last introduce Bennett's exorcism of Maxwell's demon:
conceptually  Maxwell's demon may be formalized as a computer
that:
\begin{enumerate}
  \item gets the input $ ( s , v ) $  from a device measuring both the side s from which the molecule arrives and its velocity
  \item computes a certain semaphore-function $ ( s , v ) \stackrel{p}{\rightarrow}  p[(s,v)] $ giving as output a 1 if the molecule must be
  left to pass while gives as output a 0 if the molecule must be
  stopped:

  specifically, the semaphore-function may be defined through the
  following Mathematica expression \cite{Wolfram-96}:
\begin{multline}
 p[ s_{-} \, , \, v_{-} ] \; := \; If[ s \, = \, Left \; , \;
 If[ v \leq v_{T} \, , \, 0 \, , \, 1 ] \; , \\
  If[ v > v_{T} \, , \, 0 \, , \, 1 ]]
\end{multline}
where $ v_{T} $ is a fixed threshold velocity
  \item gives the output p[(s,v)] to a suitable device that operates on the molecule in the specified way
\end{enumerate}
Both the first and the third phases  of this process, taking into
account also the involved devices, may be made in a
thermodinamically-reversible way.

As to the second step, anyway, let us observe that the
semaphore-function p is logically-irreversible and hence, by
Landauer's Principle, also thermodinamically-irreversible.

As above specified, such a thermodinamically-irreversibility may
be avoided conserving the garbage; let us, precisely, suppose,
that the demon-computer computes the
thermodynamically-reversibly-computable function $ \tilde{p} $:
\begin{multline}
 \tilde{p}[ s_{-} \, , \, v_{-} ] \; := \; If[ s \, = \, Left \; ,
 \\
 If[ v \leq v_{T} \, , \, (( s \, , \, v ) \, , \, 0) \, , \, (( s \, , \, v ) \, , \, 1) ] \; ,
 \\
  If[ v > v_{T} \, , \, (( s \, , \, v ) \, , \, 1) \, , \, (( s \, , \, v ) \, , \, 0) ]]
\end{multline}

Let us suppose to make the demon-computer operate n times on n
different molecules.

When n grows the demon, with no expenditure of work, raises the
temperature of B and lowers that of A.

But let us now analyze more carefully Clausius's  formulation of
the Second Principle: it states that no thermodynamical
transformation is possible that has as its only result the
passage of heat from a body at lower temperature to a body at
higher temperature.

In the above process the passage of heat from A to B is not the
only result: another result is the storage in the
demon-computer's memory of the n-ple of inputs $ ( ( s_{1} ,
v_{1} ) \, , \, \cdots \, , \, ( s_{n} , v_{n} ) ) $.

To make the passage of heat from A to B to become the only result
of the process we could think that the demon, at the end, erases
his memory; but this, as we have seen, cannot be done in a
thermodynamically-reversible way: such an erasure causes an
increase of entropy that may be proved to be greater than or
equal to the entropy-decrease produced by the passage of heat
from A to B.

Bennett's exorcism of Maxwell's demon, has, anyway, a far
reaching conseguence; supposed that the gas is described by the
thermodynamical ensemble  $ ( X \, , \, P ) $, let us introduce
the Bennett's entropy of P:
\begin{equation} \label{eq:Bennett's entropy}
  S_{Bennett}(P) \; :=  \;  H(P) + I(P)
\end{equation}
where:
\begin{equation} \label{eq:Shannon's entropy}
  H(P) \; := \; < \,-  \log_{2} P \, >
\end{equation}
is Shannon's entropy of the distribution P (i.e. its Gibbs'
entropy in thermodynamical language), while:
\begin{equation}
  I(P) \; := \;
  \begin{cases}
    \min \{ | x | \, : \, U(x) \, = \, P \} & \text{if $ \exists \, x  \, : \, U(x) = P $}, \\
    + \infty & \text{otherwise}.
  \end{cases}
\end{equation}
is its \textbf{prefix-algorithmic-information} (denoted simply as
\textbf{algorithmic information} form here and beyond), i.e. the
length of the shortest program computing it on the fixed Chaitin
universal computer U (demanding to \cite{Calude-02} for details we
recall that a Chaitin universal computer is a universal computer
with prefix-free halting set and the property that, up to an
input-independent additive constant, it describes algorithmically
any output in a way more concise that any other computer).

Bennett's exorcism of Mawxell's Demon implies  Bennett's Theorem
stating that the thermodynamical entropy of the ensemble $ ( X \,
, \, P ) $ is equal to to its Bennett's entropy:
\begin{equation} \label{eq:Bennett's theorem}
  S_{therm} ( P ) \; = \;  S_{Bennett}(P)  \; \neq H (P)
\end{equation}

To understand why Bennett's exorcism  implies eq.\ref{eq:Bennett's
theorem} let us consider some example:

let us suppose, for simplicity, that the initial equilibrium
probability distribution is such that the molecules have one of
only two possible velocities $ v_{L} $ and $ v_{H} $, respectively
lower and higher than the threshold velocity $ v_{T} $
\begin{equation}
   v_{L} \; < \; v_{T} \; < \; v_{H}
\end{equation}
Let us start from the case in which:
\begin{equation}
  P( v = v_{L} ) \; = \; P( v = v_{T} ) \; = \; \frac{1}{2}
\end{equation}
Supposing that the demon memorizes in the cbit $ x_{n} $ the value
of the semaphore function of the $ n^{th} $ molecule that he
observes, we have that, at the beginning, the string $ \vec{x}_{n}
\, := \, x_{1} \cdots  x_{n} $ seems to increase its length in an
algorithmically-random way, i.e. in a way such that, informally
speaking \footnote{the sign $ \approx $ of "rough-equality" may be
formalizied more precisely by the condition that  the l.r.h. grows
at least as quicker than the r.h.s. so that eq.\ref{eq:growing of
prefixes' algorithmic information} may be rigorously considered as
as a different way of stating Chaitin's condition  $ lim_{n
\rightarrow \infty}  I( \vec{x}_{n} ) \, - \, n \; = \; + \infty $
}:
\begin{equation} \label{eq:growing of prefixes' algorithmic information}
  I( \vec{x}_{n} ) \; \approx \; n
\end{equation}
As the distribution of  the molecules becomes more and more
disuniform, with the slow molecules accumulating on the left side
and the speed molecules accumulating on the right side (i.e. when
the temperature's difference among the two sides arises), the
probability distribution of $ x_{n} $ becomes more and more unfair
preferring for $ x_{n} \, = \, 0 $, so that the string $
\vec{x}_{n} $ increases its deviation from Borel-normality.

Such an increasing regularity of $ x_{n} $ corresponds to the
fact that its algorithmic-information becomes to increase more
and more slowly.

Reasoning in terms of a finite number N of molecules
\footnote{This must be considered only as an artifice to clarify
the argument, since one has to remember that the thermodynamical
limit $ N \rightarrow \infty $ has to be taken at the end; the
qualitative behaviour here described becomes, in this limit, an
asymptotic one}, after a certain number $ n_{ord} $ of
measurements made by the demon, the system reaches the state in
which all the slow molecules are on the left side of the vessel,
while all the speed molecules are on the right side; from that
point further the demon stops every molecule so that:
\begin{equation}
  x_{n} \; = \; 0 \; \; \forall n > n_{ord}
\end{equation}
At this point, in which the demon has completed its task of
lowering the \textbf{probabilistic information} of the gas so
that such a \textbf{probabilistic information} ceases to
decrease, the \textbf{algorithmic information} of the string $
\vec{x}_{n} $ ceases to increase:
\begin{equation}
  I( \vec{x}_{n} ) \; = \;   I( \vec{x}_{n_{ord}} ) \; \; \forall n > n_{ord}
\end{equation}
The whole process may, conseguentially, be seen as a transfer of
information from the gas to demon's memory in which an amount of
gas' \textbf{probabilistic information} is transferred to the
demon as \textbf{algorithmic information}.

Let us now consider the case in which the initial distribution of
molecules' velocities is unfair:
\begin{align}
  P( v = v_{L} ) & \; = \;  1 - \alpha  \\
   P( v = v_{H} ) & \; = \; \alpha
\end{align}
where $ \alpha \, \neq  \, \frac{1}{2} $.

The qualitative behaviour of the process is analogous to the
previously discussed one although the greater is the difference $
| \alpha - \frac{1}{2} | $ the littler is the amount of gas'
  \textbf{probabilistic information} converted into \textbf{algorithmic
  information}.

Now, as we have already stressed, the involved thermodynamical
process  doesn't violate the Second Principle of Thermodynamics
since the passage of heat from the low-temperatures-source A to
the high-temperature-source B is not the only result: an other
result is the memorization in demon's memory of the sequence $ \{
x_{n} \} $.

Such a memorization, that as we have seen is a transfer of
information from the gas to the demon as well as a transfer of a
portion of the overall information of the Universe from
\textbf{probabilistic} to \textbf{algorithmic} form, corresponds
to an accumulation \textbf{in algorithmic form} of useful-energy
(i.e. of energy that may be transformed in work), i.e. in an
accumulation of thermodynamical-entropy \textbf{in
algorithmic-form} that has to be counted in the Universe's
overall thermodynamical balance preventing, indeed,  the Second
Principle to be violated.

\section{Thermodynamical entropy, Statistical Mechanics and the Kolmogorovian Foundation of Information
Theory} \label{sec:Thermodynamical entropy, Statistical Mechanics
and the Kolmogorovian Foundation of Information Theory}

Despite Richard Feynman's strongly authoritative acclamation of
the Landauer-Bennett's results on Maxwell's Demons (cfr. the
section5.1.1 "Maxwell's Demon and the Thermodynamics of
Measurement" of \cite{Feynman-96}) and its appreciation by Nobel
prize awarded theoretical physicists such as Murray Gell-Mann
(cfr. e.g. the $ 15^{th} $ chapter "Time's arrows" of
\cite{Gell-Mann-94}), these, and in particular Bennett's Theorem,
are far from having being accepted by the Theoretical Physics'
community.

The objections (implicitely or explicitely) moved to Bennett's
Theorem are essentially the following:
\begin{enumerate}
  \item the Mawxell-demon's issue simply shows that the Second Law has a statistical validity
  \item the action of Mawxell's demon moves the system out of
  thermodynamical equilibrium: conseguentially the
  thermodynamical entropy ceases to be defined
  \item the interdisciplinary attitude of Algorithmic Physics is not necessary to understand Thermodynamics
\end{enumerate}
The first objection, i.e. the claim that the Second Law has only
a statistical validity, is the $ \kappa o \iota \nu \acute{\eta}
$ as far as the Mathematical-Physics' literature is concerned.

Such a claim is, anyway, false:

though Statistical Mechanics (historically pioneered by Maxwell,
Thomson and Boltzmann: cfr. e.g. the chap.3-7 of
\cite{Cercignani-98}) allows to obtain the Equilibrium
Thermodynamics of a macroscopic thermodynamical system deriving
it from a probabilistic description of its underlying microscopic
dynamics, Thermodynamics is a perfectly self-consistent physical
theory predicting the value and dynamical evolution of all the
thermodynamical observables of thermodynamical systems, (generally
not in thermodynamical equilibrium), with certainty. This occurs,
in particular, as to the Second Law of Thermodynamics stating that
in any thermodynamical cycle of any isolated thermodynamical
system (generally not in thermodynamical equilibrium) one has with
certainty that:
\begin{equation} \label{eq:second principle of thermodynamics}
 \oint \frac{ \delta Q}{ T} \; \leq \; 0
\end{equation}
where $ \delta Q $ is heat's amount absorbed by the system while T
is its temperature.

The source of the erroneous claim that Maxwell-demon's issue
simply shows the statistical validity of the Second Law may be
understood in terms of the following words by Joel Lebowitz:
\begin{center}
 \textit{"The various ensembles commonly used in statistical mechanics are
 to be thought of as nothing more than mathematical tools for
 describing behaviour which is practically the same for "almost
 all" individual macroscopic systems in the ensemble. While these
 tools can be very useful and some theorems that are proven about
 them are very beautiful they must not be confused with the real thing going in a single system.
 To do that is to commit the scientific  equivalent of idolatry,
 i.e. substituting representative images for reality"; cited from the Introduction of \cite{Lebowitz-95}}
\end{center}
Such an idolatric attitude for which a thermodynamical system is
confused with its modellization trough Statistical Mechanics is,
indeed, a typical mental  attitude of some Mathematical
Physicists that had often induced even authoritative scientists
to assert trivially erroneous statements of Thermodynamics; this
is the case, for example, of Giovanni Gallavotti's analysis of
brownian motors that, misunderstanding the celebrated analysis by
Richard Feynman of a "ratchet and pawl heat engine":
\begin{center}
 \textit{"Let us try to invent a device which will violate the Second Law of Thermodynamics, that is a gadget which will generate
 work from a heat reservoir with everything at the same temperature. Let us say we have a box of gas at a certain temperature, and inside
 there is a an axes with vanes in it. $\cdots$. Because of the bombardments of gas molecules on the vane, the vane oscillates and jiggles.
 All we have to do is to hook into the other end of the axle a wheel which can turn only one way- the ratchet and pawl. Then when the shaft tries to jiggle one way, it will not turn, and when it jiggles
 the other, it will turn. Then the wheel will slowly turn, and perhaps we might even tie a flea onto a string hanging from a drum on the shaft, and lift
 the flea! Now let us ask if this is possible. According to Carnot's hypothesis, it is impossible. But is we just look at it we see, prima facie, that it seems quite possible. So we must look more closely"; cited from the chapter 46 of \cite{Feynman-63a}}
\end{center}
streghtens Jean Perrin's restatement of the idolatric claim that
the Second Law has only a statistical validity:
\begin{center}
  \textit{"But is must be remembered $ \cdots $ that the brownian
  movement, which is a fact beyond dispute, provides an
  experimental proof (deduced from the molecular agitation
  hypothesis) by which of means Maxwell, Gibbs and Boltzmann
  robbed \emph{Carnot's Principle} of its claim to rank as an absolute truth and reduced it to the
  mere expression of a very high probability"; cited from the
  $51^{th}$ section \emph{"The brownian movement and Carnot's
  Principle"} of \cite{Perrin-90}}
\end{center}
claiming that:
\begin{center}
  \textit{"It is important to keep in mind that here we are somewhat stretching the validity of thermodynamics laws: the above machines are
  very idealized objects, like the daemon. They cannot be realized in any practical way; one can arrange them to perform one cycle, $ \cdots $ but what one needs to violate the second law is the possibility
  of performing as many energy producing cycles as  required.Otherwise their existence "only" proves  that the second law has only a statistical validity, a fact
  that had been well establishes with the work of Boltzmann.
  In fact an accurate analysis of the actual possibility of building walls semipermeable to colloids and of exhibiting macroscopic violation
  of the second principle runs into grave difficulties: it is not possible to realize a perpetual motion of the second kind by using the properties of Brownian motion.
  It is possible to obtain a single violation of Carnot's law (or of a few of them) of the type described by Perrin, but as time
  elapses and the machine is left running, isolated and subject to physical laws with no daemon or other extraterrestrial being intervening (or performing work accounted for), the violations (i.e. the energy produced per cycle) vanish because the cycle will be necessarily performed as many times in one direction (apparently violating Carnot's principle and producing work) as in the opposite direction (using it).
  This is explained in an analyis on Feynman, see \cite{Feynman-63a} chapter 46, where the semi-permeable wall is replaced by a wheel with an anchor mechanism, a "ratchet and a pawl", allowing it to rotate only in one direction
  under the impulses communicated by the colloidal particle collisions with the valves of a second wheel rigidly bound to the same axis. Feynman's analysis is really beatiful, and remarkable as an example of how one can still say something interesting on perpetual motion. It also brings important insight into the related so-called "reversibility paradox" (that microscopic dynamics generates an irreversible macroscopic world)."; cited from the section8.1 "Brownian motion and Einstein's Theory" of \cite{Gallavotti-99}}
\end{center}
i.e. that a brownian motor can violate Carnot's Law for a few
cycles, a thing that, if it was true (that unfortunately this is
not the case is shown, for example,  in \cite{Astumian-97}), would
have allowed Gallavotti to definitively resolve the energetic
problem of the World saving it from the slavery of oil.

\smallskip

The second objection (implicitely or explicitely) moved to
Bennett's Theorem, namely that since the action of Mawxell's demon
moves the system out of thermodynamical equilibrium  the
thermodynamical entropy ceases to be defined, is based again on
the idolatric attitude of making confusion between a
thermodynamical system and its modellization through Statistical
Mechanics denounced by Lebowitz:

the  fact that there doesn't exist a universally accepted notion
of entropy in Nonequilibrium Statistical Mechanics is
consequentually seen a synonimous of the false statement that the
notion of thermodynamical entropy of a thermodynamical system not
in equilibrium is not defined.

Such a confusion appears, for example, in the following passage
by Gallavotti:
\begin{center}
  \textit{"One of the key notions in equilibrium statistical mechanics is
  that of \emph{entropy}; its extension is surprisingly difficult,
  assuming that it really can be extended. In fact we expect that,
  in a system that reaches under forcing a stationary state,
  entropy is produced at a costant rate, so that there is no way
  of defining an entropy value for the system, except perhaps by
  saying that its entropy is $ - \infty $. Although one should
  keep in mind that there is no universally accepted notion of
  entropy in systems out of equilibrium, even when in a stationary state,
  we shall take the attitude that in a stationary state only the
  entropy \textbf{creation rate} is defined: the system entropy
  decreases indefinitely, but at at costant rate. Defining "entropy" and "entropy production" should
  be considered an open problem"; cited from the section9.7
  "Entropy Generation. Time Reversibility and Fluctuation Theorem.
  Experimental Tests of the Chaotic Hypothesis" of \cite{Gallavotti-99}}
\end{center}
or in the following passage by Olivier Penrose:
\begin{center}
  \textit{"Even in thermodynamics, where entropy is defined only for
  equilibrium states, the definition of entropy can depend on what
  problem we are interested in and on what experimental techniques
  are available"; cited from \cite{Penrose-79}}
\end{center}
that is implicitly a kind of self-criticism as to the following
analysis of Mawxwell's-demon:
\begin{center}
\textit{ "The large number of distinct observational states that the
 Maxwell demon must have in order to make significant entropy
 reductions possible may be thought of as a large memory capacity
 in which the demon stores the information about the system which
 he acquires as he works reducing the entropy. As soon as the
 demon's memory is completelly filled, however, $ \cdots $ he can achieve no further
 reduction of the Boltzmann entropy. He gains nothing for example, by deliberately forgetting or erasing some of his stored
 information in order to make more memory capacity available; for
 the erasure being a setting process, itself increases the entropy
 by an amount of at least as great as the entropy decrease made
 possible by the newly available memory capacity"; cited from \cite{Penrose-70}}
\end{center}
in which, as it has been observed by Harvey S. Leff and Andrew F.
Rex in the $ 1^{th} $-chapter "Overview" of their wonderful book
\cite{Leff-Rex-90}, Olivier Penrose arrived very near to the
right Bennett's exorcism, though lacking to make the final
intellectual step to understand that erasure is the
\emph{fundamental} act that saves Mawxell's demon.

Let us now explicitely show how Lebowitz's remark against
idolatry allows to confute the first objection to Bennett's
Theorem, namely that the Second Law of Thermodynamics has only
statistical validity: let us analyze, at this purpose, the
following pass in which Maxwell himself explains what he wanted to
show through the introduction of his demon:
\begin{center}
  \textit{"This is only one of the instances in which conclusions which we have drawn from our experience
  of bodies consisting of an immense number of molecules may be
  found not to be applicable to the more delicate observations and
  experiments which we may suppose made by one who can perceive
  and handle the individual molecules which we deal with only in
  large masses.
  In dealing with masses of matter, while we do not perceive the
  individual molecules, we are compelled to adopt what I have
  described as the statistical method of calculation, and to
  abandon the strict dynamical method, in which we follow every
  motion by the calculus.
  It would be interesting to enquire how far those ideas about the
  nature and method of science which have been derived from
  examples of scientific investigation in which the dynamical
  method is followed are applicable to our actual knowledge of
  concrete things, which, as we have seen, is of an essentially
  statistical nature, because no one has yet discovered any
  practical method of tracing the path of a molecule, or of identifying
  it at different times."; cited
from the last but one section "Limitation of the Second Law of
Thermodynamics" of the $22^{th}$ chapter of \cite{Maxwell-01}}
\end{center}
and the following pass by Thomson (later Lord Kelvin):
\begin{center}
  \textit{"'Dissipation of Energy' follows in nature from the fortuitous
  concourse of atoms. The lost motivity is essentially not
  restorable otherwise than by an agency dealing with individual
  atoms"; cited from \cite{Thomson-90}}
\end{center}
They don't say that the Second Principle of Thermodynamics have
only a statistical validity, but a different thing: that the
usual link existing between such a principle (that, not falling in
the idolatry denounced by Lebowitz, one have to remember to have
an its own validity in Thermodynamics) and Statistical Mechanics
have to be modified as soon as entitities able to handle
individual molecules are involved.

\smallskip

Having followed Lebowitz's advise of not falling into the
idolatric attitude of making confusion among a physical
thermodynamical system and its modellization through Statistical
Mechanics and, hence, preserving us from the error of confusing
the difficulties involved in the definition of entropy in
Nonequilibrium Statistical Mechanics from the difficulties
involved in defining entropy in Nonequilibrium Thermodynamics, let
us briefly recall these latter:

given a thermodynamical system made of N different species, the
thermodynamical entropy of a thermodynamical state X is defined
as:
\begin{equation} \label{eq:thermodynamical entropy}
  S_{therm} (X) \; := \; \int_{REV}^{X} \frac{ \delta Q }{T}
\end{equation}
where the integral is over a thermodynamically-reversible
tranformation starting in a fixed reference thermodynamical state
O (to be ultimatively fixed by the Third Law of Thermodynamics
requiring that $ \lim_{T \rightarrow 0} S_{therm} (X) \, = \, 0
$)  and ending in the state X.

If X is a state of thermodynamical equilibrium the thermodynamical
entropy may be expressed as a function of the internal energy U,
of the volume V and of the number of moles of each contributing
specie $ N_{k} $:
\begin{multline} \label{eq:thermodynamical entropy as function of
other state's variables at equilibrium} X \text{ equilibrium
state} \; \Rightarrow \\
 S_{therm} (X) \; = \; S_{therm}[ U(X) \, , \, V(X) \, , \,  N_{k}
 (X) ]
\end{multline}
If X is not a state of thermodynamical equilibrium, anyway, its
thermodynamical entropy cannot be  expressed anymore as a function
of the internal energy U, of the volume V and of the number of
moles of each contributing specie $ N_{k} $:
\begin{multline} \label{eq:thermodynamical entropy not  function of other state's variables at nonequilibrium}
X \text{nonequilibrium state} \; \Rightarrow \\
 S_{therm} (X) \; \neq \; S_{therm}[ U(X) \, , \, V(X) \, , \,  N_{k}
 (X) ]
\end{multline}
This fact is often erroneously expressed as the claim that
thermodynamical entropy is not defined out of equilibrium: this is
simply false, since the the operational definition of $ S_{therm}
(X) $ through eq.\ref{eq:thermodynamical entropy} continues to
hold.

Simply one has, denoting with lower case letters the (intensive)
densities of (extensive) quantities, that the equation
eq.\ref{eq:thermodynamical entropy as function of other state's
variables at equilibrium} of Classical Thermodynamics must be
generalized by its expression in Generalized Thermodynamics,
having the form \cite{Jou-Casas-Vasquez-Lebon-01}:
\begin{multline} \label{eq:thermodynamical entropy's density as function of other state's variables at nonequilibrium}
 s_{therm} (X,t ) \; = \; s_{therm}[ u(X,t) \, , \, v(X,t) \, , \,  n_{k}
 (X,t) \, , \\ \nabla u(X,t) \, , \, \nabla v(X,t) \, , \,  \nabla
 n_{k}
 (X,t) \, , \\
  \,\nabla^{2} u(X,t) \, , \,   \nabla^{2} V(X,t) \, , \, \nabla^{2}  n_{k}
 (X,t), \, \cdots \,  ]
\end{multline}
and reducing to eq.\ref{eq:thermodynamical entropy as function of
other state's variables at equilibrium} in the equilibrium case.

Under conditions explicitely formalizable, furthermore, the Local
Equilibrium Condition, stating that the local and istantaneous
relations between the thermal and mechanical properties of a
physical system are the same as for a uniform system at
equilibrium, holds. In this case eq.\ref{eq:thermodynamical
entropy's density as   function of other state's variables at
nonequilibrium} reduces to:
\begin{equation} \label{eq:thermodynamical entropy's density under Local Equilibrium Condition}
 s_{therm} (X,t ) \; = \; s_{therm}[ u(X,t) \, , \, v(X,t) \, , \,  n_{k}
 (X,t) ]
\end{equation}
i.e.:
\begin{equation}
  ds  \; = \; ( \frac{ \partial s  }{ \partial u  } )_{v , n_{k}}
  \, + \, ( \frac{ \partial s  }{ \partial v  } )_{u , n_{k}}
  \, + \, \sum_{k=1}^{N} ( \frac{ \partial s  }{ \partial n_{k}  } )_{u , v ,
  n_{k'}} \; (  \, for \,  k' \neq k )
\end{equation}
Consequentially, if the one-parameter family of nonequilibrium
thermodynamical states $ X_{t} $  satisfy the Local-Equilibrium
Condition, one has that the temperature in the point $ \vec{x} $
of the system at time t may be simply expressed as:
\begin{equation}
  T( X_{t} , \vec{x} , t ) \; = \; [( \frac{ \partial s  }{ \partial u  } )_{v ,
  n_{k}}]^{- 1}
\end{equation}
Returning at last to our Maxwell's demon, let us observe  that its
way of taking the whole thermodynamical system out of the
thermodynamical equilibrium satisfies the conditions under which
the Local-Equilibrium Condition holds.

\smallskip

The third objection moved (implicitely or explicitely) to
Bennett's Theorem, namely that the intedisciplinary attitude of
Algorithmic Physics is not necessary to understand
Thermodynamics, is certainly the subtler one.

To analyze it, let us observe that the involved thermodynamical
system is the compound system Gas + Demon.

As a mechanical system such a compound system is a classical
dynamical system  $ ( X_{compound} \, , \, H_{compound} )$ with
phase space:
\begin{equation}
  X_{compound} \;  \; := \;  X_{Gas} \, \bigcup \, X_{Demon}
\end{equation}
and hamiltonian:
\begin{equation}
  H_{compound} \; := \; H_{Gas} \, + \,  H_{Demon} \, + \,
  H_{interaction}
\end{equation}
where $ H_{Gas} \in C^{\infty} ( X_{Gas} ) $, $ H_{Demon} \in
C^{\infty} ( X_{Demon} ) $, and $ H_{interaction} \in C^{\infty} (
X_{compound} ) $.

The mechanical description of the whole process is defined by the
Hamiltonian flow $ T_{t} : X_{compound} \rightarrow X_{compound}
$ induced by Hamilton's equation:
\begin{equation} \label{eq:Hamilton's equation}
  \frac{d x}{d t} \; = \; \{ H \, , \, x \}
\end{equation}
associating to any initial state $ x_{compound}^{(IN)} \, = \, (
x_{Gas}^{(IN)} , x_{Demon}^{(IN)} ) \in X_{compound} $ the final
state $ x_{OUT}  \, := \, \lim_{ t \rightarrow \infty} T_{t}
x_{IN} $.

The strategy of Statistical Mechanics would consist in deriving
the macroscopic thermodynamical variables of the thermodynamical
system Gas+Demon  as properly-defined  functions of a suitable
statistical ensemble $ ( X_{compound} \, , \, P_{compound} ) $.

The ensemble $ ( X \, , \, P) $ involved in the formulation of
Bennett's Theorem, instead, doesn't take into account  the demon:
as we saw in the last section, it is the equilibrium statistical
ensemble $ ( X_{gas} \, , \, P_{eq} ) $ that Statistical
Mechanics would associate to the dynamical system $ ( X_{Gas} \,
, \, H_{Gas} ) $, the underlying reason for that deriving
substantially from the Algorithmic Physics' attitude, as we will
now explain.

Algorithmic Physics is, by definition, that discipline analyzing
physical processes looking at them as computational processes
according to the following correspondence's table:

\smallskip

\begin{tabular}{|c|c|}
  PHYSICAL PROCESS   &  COMPUTATIONAL PROCESS  \\ \hline
  initial state &  input \\
  dynamical evolution &  computation \\
  final state  &  output \\ \hline
\end{tabular}

\smallskip

and conseguentially applying the conceptual instruments of
Computation's Theory.

Essentially owing to the overwhelming "new age" folklore by which
it has been popularized in the divulgative literature, the
interdisciplinary nature of Algorithmic Physics is looked by many
theoretical and mathematical physicists with great mistrust; as a
consequence, also the  beautiful and serious insight it has
produced, such as the investigations concerning the foundations
of Computational Physics (i.e. of the discipline studying the
computer-simulation of physical systems)  such as Stephen
Wolfram's notion of computational irreducibility (i.e. the
situation in which the faster way of predicting the final state
of a dynamical system of known laws-of motion is to simulate its
whole dynamical evolution and to see what happens at the end) or
his analyses concerning the rule of Undecidability in Physics
\cite{Wolfram-85} \footnote{such as Wolfram's professional path
has been characterized by his departure from academia in order of
constituting  Wolfram Research Inc. producing Mathematica
\cite{Wolfram-96}, his intellectual path has been characterized by
an analogous non-conformism that led him to develop his seminal
ideas culminating  into a personal conception of randomness and
complexity radically different by that of Algorithmic Information
Theory \cite{Bennett-88} and based on what he calls the Principle
of Computational Equivalence;  the inter-relations existing among
such a  viewpoint and Algorithmic Information Theory is analyzed
in the first four sections "Introduction", "What Perception and
Analysis Do", "Defining the Notion of Randomness", "Defining
Complexity" of the $10^{th}$ chapter "Processes of Perception and
Analysis" and in the $12^{th}$ chapter "The Principle of
Computational Equivalence" of \cite{Wolfram-02}, in the final
section "Afterthoughts..."  of the $7^{th}$ chapter "Mathematics
in the Third Millenium" of \cite{Chaitin-99} and  in the final
section "Final remarks" of the $4^{th}$ part "Future Work" of
\cite{Chaitin-01}} concretized by the work of Chris Moore and
many others \cite{Svozil-93}, \cite{Moore-98}, are ignored.

The approach underlying Bennett's Theorem is a partial application
of the Algorithmic Physics' approach in which not the whole
hamiltonian flow $ T_{t} : X_{compound} \mapsto X_{compound} $ is
seen as a computational process, but only its restriction as to
the Demon $ T_{t} |_{ X_{Demon} } :  X_{Demon} \mapsto  X_{Demon}
$.

The dynamical evolution  of the gas $ T_{t} |_{ X_{Gas} } :
X_{Gas} \mapsto  X_{Gas} $ continues to be described through
Mechanics, i.e. owing to the enormous number of involved degrees
of freedom, through Statistical Mechanics.

The third objection to Bennett's theorem is based on the
observation that such an (hybrid) recourse to Algorithmic Physics
is, at last, completelly avoidable:

why, for particular compound systems,  should one to give up the
usual, traditional approach of Statistical Mechanics to derive
the thermodynamical entropy in the usual way from the partition
function of a Gibbs's ensemble for the dynamical system $ (
X_{compound} \, , \, H_{compound} )$?

And which should be exactly these particular compound systems?

A minimal answer to the last question is immediate: those
particular compound systems in which $ X_{Demon} $ and $ H_{Demon}
$ are such to result in the  scattering pattern that, looking at
the demon with the eyes of Algorithmic Physics, corresponds to the
computational-process of computing the semaphore function p and
making to pass or to reflect the molecule correspondigly as
described in the last section; as we will see in the next section,
anyway, such a class of particular compound systems may be
considerably enlarged through a suitable characterization of the
notion of an \textbf{intelligent system}.

As to the former question, namely why for these compound systems
one should indeed to give up the usual Statistical Mechanics'
approach, the answer is: simply because it is simpler.

The third objection to Bennett's Theorem is, with this regard,
correct: there is no necessity of adopting the hybrid
Algorithmic-Physics'approach on which Bennett's Theorem is based
on:

simply,  given an arbitrary many-body physical system like $
X_{Gas} $, whenever its interaction  with another physical systems
gives rise to a scattering-cross-section $ \frac{d \sigma}{d
\Omega } $ of the particular kind specified above, the usual
Statistical Mechanics' approach, though still perfectly
applicable, is not the simpler approach since it doesn't catch
the particular  structural peculiarity of the analyzed system,
structural peculiarity that allows an alternative,  more concise,
explication that, according to Occam's Razor, have conseguentially
to be preferred.

Such a passage from a purely probabilistic approach to a hybrid
mix of two approaches, the probabilistic and the algorithmic one,
reflects itself in the the link between Thermodynamics and
Information Theory:

in terms of the three different approaches to the definition of
information introduced by A.N. Kolmogorov in his fundamental
papers on the Foundation of Information Theory
\cite{Shiryayev-94} such a passage is exactly a passage from an
interpretation of thermodynamical entropy in terms of the
probabilistic approach alone to an interpretation of
thermodynamical entropy as a hybrid mix of the probabilistic and
the algorithmic approaches.

\section{Zurek's Theorem on Maxwell's Demons in Quantum Thermodynamics} \label{sec:Bennett's Theorem on Maxwell's Demons in Quantum Thermodynamics}
Bennett's work on Maxwell's demon has been generalized by
Wojciech Hubert Zurek  in many respects \cite{Zurek-89},
\cite{Zurek-90a}, \cite{Zurek-90b}, \cite{Zurek-99}.

The first point analyzed by Zurek concerns the characterization
of the  particular structural peculiarity of the dynamical system
$ ( X_{compound} \, , \, H_{compound} ) $ under which the mix
probabilistic + algorithmic approach underlying Bennett's Theorem
may be applied:

up to this point we have assumed that Maxwell's demon acts on a
particular molecule in a  very particular way: if the molecule
arrives from the left side the demon makes it to pass unaltered if
and only if its velocity is less or equal to a given
threshold-velocity, acting in the opposite way if the molecule
arrives from the right side.

Such a behaviour of the demon, as it was first observed  by Leo
Szilard in his basic 1929's paper \cite{Szilard-90}, appears as a
kind of \emph{intelligence}.

While, taken too literally, Szilard's paper was unfortunately
also the source of many confusionary speculations concerning the
contribution of the Subject (or the Cartesian Cogito in more
philosophically palatable terms) to the Object's thermodynamical
entropy, sometimes appealing to the wrong claim that Subject's
measurements are necessary thermodynamically-irreversible
processes (whose falsity, as we have seen, may be directly
derived by Landauer's Principle; in the quantum case we are going
to introduce, anyway, it was time before directly derived by
Yakir Aharonov, Peter Bergmann and Joel Lebowitz
\cite{Aharonov-Bergmann-Lebowitz-83}), it had the great merit of
intuitively suggesting the essential structural peculiarity of
Maxwell's demon, allowing Zurek to generalize Bennett's Theorem
starting from the following questions:
\begin{enumerate}
  \item which is exactly the kind of intelligence showed by
  Maxwell's demon?
  \item can Bennett's Theorem to be generalized to systems having the
  same kind of intelligence?
\end{enumerate}
Observing that, according to the way we characterized it in the
previous section, the structural peculiarity of $ X_{demon} $ and
$ H_{demon} $ is to give rise to a scattering cross-section $
\frac{d \sigma}{d \Omega } $ that, from the point of view of
Algorithmic Physics, corresponds to the prescribed algorithm of
leaving to pass or reflecting elastically a molecule according to
the value of a computed semaphore-binary predicate p(s,v), one
could think that Bennett's Theorem might be generalized to any
situation in which $ X_{demon} $ and $ H_{demon} $ give rise to a
scattering cross-section $ \frac{d \sigma}{d \Omega } $ that,
from the point of view of Algorithmic Physics, corresponds to an
algorithm leaving to pass or reflecting elastically a molecule
according to a computed arbitrary binary predicate.

But one immediately realizes that such a generalization is wrong
in that not every chosen semaphore-function corresponds to a
resulting behaviour that seems to be intelligent.

Indeed, Szilard tells us, the particular "intelligence" of the
semaphore-predicate p derives from the fact that the resulting
algorithm performed by the demon acts on each molecule in order of
lowering the probabilistic information of the gas: we are tempted
to say that it acts in a clever way exactly since his behaviour
seems to be teleological, finalized to the objective of taking the
gas in a more ordered state.

Let us observe, finally, that such an ordering-process made by the
demon acts necessarily out of thermodynamical equilibrium, since
its "intelligence" is accomplished precisely by creating a
"clever" disuniformity in the spatial distribution of a
thermodynamical variable (in this case the temperature).

We are conseguentially led to the following generalization of
Bennett's theorem:

the thermodynamical entropy of a classical many-body system $ (
X_{m.b} , H_{m.b} ) $:
\begin{itemize}
  \item preliminary prepared in a state of thermodynamical
equilibrium described, in Classical Statistical Mechanics, by the
statistical ensemble $ ( X_{m.b} , P_{eq} ) $
  \item in a second time made to interact with an other physical system  $ ( X_{int} \, , \,
H_{int} ) $ that is \textbf{intelligent}:
\end{itemize}
may be expressed as:
\begin{equation}\label{eq:generalized Bennett's theorem}
   S_{therm} ( P_{eq} ) \, = \, I_{prob} ( P_{eq} ) \, + \,  I_{alg} ( P_{eq} )
\end{equation}
where  $ I_{prob} ( P_{eq} ) $ and  $ I_{alg} ( P_{eq} ) $ are,
respectively, the probabilistic information and the algorithmic
information of the ensemble $ ( X_{m.b} , P_{eq} ) $,  where the
intelligence-condition of $ ( X_{int} \, , \, H_{int} ) $ is
defined  in the following way:
\begin{enumerate}
  \item the scattering cross-section $ \frac{d \sigma}{d \Omega }
  $, seen from the point of view of Algorithmic Physics,
  corresponds to a deterministic algorithm  f acting on a single molecule
  \item in a way that takes  the many-body system  out of
  thermodynamical equilibrium
  \item reducing its probabilistic information
\end{enumerate}
Such a definition of an \textbf{intelligent system} is, indeed, a
strenghtening of  Gell-Mann's notion of \textbf{information
gathering and using system (IGUS)} defined as a \textbf{complex
adaptive system} able to make observations (cfr.
\cite{Gell-Mann-94} and the $12^{th} $ section of
\cite{Gell-Mann-Hartle-90}) that, contrary to these notions
difficult to formalize, has a precise mathematical meaning.

A part from having generalized it to a strongly larger class of
intelligent systems, Zurek's main extension of Bennett's work
concerns its extension to the quantum domain:

Zurek's theorem \footnote{I would like to advise the reader that
my presentation of Zurek's theorem differs slightly from Zurek's
own ideas for which we strongly demand to the previously cited
original Zurek's papers}, the quantum analogue of Bennett's
theorem, states that the thermodynamical entropy of a quantum
many-body system $ ( {\mathcal{H}}_{m.b} , \hat{H}_{m.b} ) $:
\begin{itemize}
  \item preliminary prepared in a state of thermodynamical
equilibrium described, in Quantum Statistical Mechanics, by the
quantum statistical ensemble $ ( {\mathcal{H}}_{m.b} , \rho_{eq} )
$
  \item in a second time made to interact with an other physical system  $ ( {\mathcal{H}}_{int} \, , \,
\hat{H}_{int} ) $ that is \textbf{intelligent}:
\end{itemize}
may be expressed as:
\begin{equation}\label{eq:Zurek's theorem}
   S_{therm} ( \rho_{eq} ) \, = \, I_{prob} ( \rho_{eq} ) \, + \,  I_{alg} ( \rho_{eq} )
\end{equation}
where  $ I_{prob} ( \rho_{eq} ) $ is the quantum probabilistic
infomation of the density operator $ \rho_{eq} $, namely its Von
Neumann's entropy:
\begin{equation}
  I_{prob} ( \rho_{eq} ) \; := \; - \, Tr  \rho_{eq} \log \rho_{eq}
\end{equation}
while  $ I_{alg} ( \rho_{eq} ) $ is the quantum algorithmic
information of the ensemble $ ( {\mathcal{H}}_{m.b} , \rho_{eq} )
$ \footnote{Zurek claims that the assumption of the
Church-Turing's Thesis eliminates any dependence from the
particular universal computer U adopted by (or better
constituting) the demon. He, in particular, claims that, by the
Church-Turing's Thesis, it doesn't matter if U is a
\textbf{classical computer} or a \textbf{quantum computer}; so he
substantially claims that Quantum Algorithmic Information Theory
collapses to Classical Algorithmic information Theory, an
arbitrary assumption for whose discussion I demand to
\cite{Segre-02}. I will therefore assume, from here and beyond,
that the computer U in eq.\ref{def:Zurek's entropy of a density
operator} is a Universal Quantum Computer. But let then observe
that, in this way, one implicitly assumes that the quantum
algorithmic information of a quantum state must be defined in
terms of classical-descriptions of such a state, as claimed by
Svozil \cite{Svozil-96} and Vitanyi \cite{Vitanyi-01}, and not in
terms of quantum descriptions as it is claimed by Berthiaume, Van
Dam and Laplante \cite{Berthiaume-van-Dam-Laplante-00}, an issue,
this one, about which  I demand once more to \cite{Segre-02}},
namely:
\begin{equation} \label{def:Zurek's entropy of a density operator}
  I( \rho_{eq}) \; := \;
  \begin{cases}
    \min \{ | x | \, : \, U(x) \, = \, \rho_{eq} \} & \text{if $ \exists \, x  \, : \, U(x) = \rho_{eq} $}, \\
    + \infty & \text{otherwise}.
  \end{cases}
\end{equation}
i.e. the length of the shortest program computing it on the fixed
Chaitin quantum universal computer U, where the
intelligence-condition of
 the quantum system $ ( {\mathcal{H}}_{int} \, , \,
\hat{H}_{int} ) $ is defined exactly as in the classical case
through the following conditions:
\begin{enumerate}
  \item the scattering cross-section $ \frac{d \sigma}{d \Omega }
  $, seen from the point of view of Algorithmic Physics,
  corresponds to an  algorithm  f acting on a single molecule
  \item in a way that takes  the many-body system  out of
  thermodynamical equilibrium
  \item reducing its probabilistic information
\end{enumerate}
It is important to remark, at this point, that such a definition
of intelligence, applied in the quantum domain, is indeed subtler,
owing to the entanglement's phenomenon between the quantum
many-body system $ ( {\mathcal{H}}_{m.b} , \hat{H}_{m.b} ) $ and
$ ( {\mathcal{H}}_{int} \, , \, \hat{H}_{int} ) $ having no
classical analogue, that is itself used by $ (
{\mathcal{H}}_{int} \, , \, \hat{H}_{int} ) $ as to realize its
teleological action.

\section{The thermodynamical cost of erasing the membranes'
memory of Peres' engine} \label{sec:the thermodynamical cost of
erasing the membranes' memory of Peres' engine}

In the previous sections we have introduced all the ingredients
required to present the contribution of this paper, namely the
confutation of the claim, presented by  Asher Peres in
\cite{Peres-90} as well as in the $ 9^{th} $ chapter of his
wonderful book \cite{Peres-95},  that the Theorem of
Indistinguishability for nonorthogonal states is necessary in
order of preserving the Second Law of Thermodynamics.

Peres's argument is based on the assumption that the
\textbf{thermodynamical-entropy} of a quantum system is described
by \textbf{Von Neumann's entropy}, assumption that he deeply
analyzes explicitly reporting the celebrated original calculus by
which Von Neumann, in the section5.2 of \cite{Von-Neumann-83},
computed the thermodynamical entropy of a quantum mixture $ \{
p_{i} \, , \, | \phi_{i} > < \phi_{i} | \}_{i=1}^{n} $  as if each
$| \phi_{i} > < \phi_{i} | $ was a specie of ideal gas enclosed in
a large impenetrable box, and inferring that the thermodynamical
mixing entropy of the different species is $ I_{prob} ( \sum_{i}
p_{i} | \phi_{i} > < \phi_{i} | ) $.

Peres reviews Von Neumann's procedure in the following way:
\begin{center}
  \textit{"It also assumes the existence of semipermeable membranes which can be used to perform quantum tests.
  \textbf{These membranes separate orthogonal states with perfect efficency}. The fundamental problem here is whether it is
  legitimate to treat \textbf{quantum states} in the same way as varieties of classical ideal gases.
  This issue was clarified by Einstein in the early days of the "old" quantum theory as follows: consider an ensemble of quantum
  systems, each one enclosed in a large impenetrable box, so as to prevent any interaction between them.
  These boxes are enclosed in an even larger container, where they behave \textbf{as an ideal gas}, because each box is so  massive
  that classical mechanics is valid for its motion ($\cdots $). The container itself has ideal walls and pistons which may be, according to our needs,
  perfectly conducting, or perfectly insulating, or with properties equivalent to those of semipermeable membranes.
  \textbf{The latter are endowed with automatic devices able to peak inside the boxes and to test the state} of the quantum system enclosed therein." (from the section9.3 of \cite{Peres-95})}
\end{center}
There is a point, anyway, of this review in which, deliberately,
Peres moves away from Von Neumann's original treatment:

he doesn't assume that \textbf{the membranes separate
nonorthogonal states with perfect efficiency} as, instead, Von
Neumann does:
\begin{center}
  \textit{"Each system $ s_{1} , \cdots ,  s_{n} $ is confined in a box $ K_{1} , \cdots ,  K_{n} $ \textbf{whose walls are impenetrable to all transmission effects} -- which is possible for this system because
  of the lack of interaction" (from the section5.2 of \cite{Von-Neumann-83})}
\end{center}

The reason why Peres, contrary to Von Neumann, doesn't make such
an assumption is that, according to him, this would imply a
violation of the Second Law of Thermodynamics; his argument is
the following: if semi-permeable membranes which unambiguously
distinguish non-orthogonal states were possible, one could use
them to realize the following cyclic thermodynamical
transformation for a mixture of two species of  1-qubit's states,
the $ | 0 > < 0 |$-specie  and the $ \frac{1}{2} ( | 0 >  + | 1 >
) ( < 0 | + < 1 | ) \} $- specie, both with the same
concentration $ \frac{1}{2}$
\begin{itemize}
  \item in the initial state the two species occupy two chambers with equal volumes, with the  $ | 0 > < 0 | $- specie occupying the right-half of the left-half of the vessel and the  $ \frac{1}{2} ( | 0 >  + | 1 > ) ( < 0 | + < 1 | ) \}  $- specie occupying the
  left-half of the right-half of the vessel
  \item the first step of the process is an isothermal expansion by which the  $ | 0 > < 0 | $- specie occupies all the left-half of the vessel while the $ \frac{1}{2} ( | 0 >  + | 1 > ) ( < 0 | + < 1 | )  $-specie occupies all the right-half of the vessel; this expansion supplies an amount
  of work:
\begin{equation}
  \Delta L _{1} \; = \; +  n T \ln 2
\end{equation}
T being the temperature of the reservoir.
  \item at this stage the impenetrable partitions separating the two species are replaced by the "magic"-semi-permeable membranes having the ability of distinguish non-orthogonal states;
precisely one of them is transparent to the $ | 0 > < 0 |
$-specie and reflect the $ \frac{1}{2} ( | 0 >  + | 1 > ) ( < 0 |
+ < 1 | ) \}  $-specie while the other membrane has the opposite
properties; then, by a double frictionless piston, it is possible
to bring the engine, without expenditure of work or heat
transfer, to a state in which all the two species occupy with
the  same concentration only the left-hand of the vessel, the
right-hand of the vessel  remaining  empty; we can represent
mathematically the state of affairs of the system by the
following decomposition:
\begin{align}
   {\mathcal{E}}_{1} & \; := \;  \{ ( \frac{1}{2} , | 0 > < 0 |  ) \, , \, ( \frac{1}{2} , \frac{1}{2} ( | 0 >  + | 1 > ) ( < 0 | + < 1 | )  )  \}  \\
  \rho  & \; := \; \begin{pmatrix}
    \frac{3}{4} & \frac{1}{4} \\
    \frac{1}{4} & \frac{1}{4} \
  \end{pmatrix}
\end{align}
  \item since the state of the mixture-of-species is completelly determined  by $ \rho $, and not by a particular its decomposition, to represent the actual state of affairs by  $ {\mathcal{E}} $ or by the Schatten's decomposition of $ \rho $:
\begin{align}
  {\mathcal{E}}_{1} & \; \; := \; \{ ( \rho_{-}  , | e_{-} > < e_{-} | ) \, , \, ( \rho_{+}  , | e_{+} > < e_{+} | ) \}  \\
  \rho_{\pm} & \;  := \;  \frac{1}{4} ( 2 \pm \sqrt{2} ) \\
   | e_{\pm} >  & \; := \; ( 1 \pm \sqrt{2} ) ( | 0 > \, + \, | 1 > )
\end{align}
is absolutely equivalent
  \item let us now replace the two "magic" membranes with ordinary membranes able to distinguish only orthogonal species;  since the $ | e_{-} > < e_{-} | $-specie and the $ | e_{+} > < e_{-} | $-specie are orthogonal, the reversible
diffusion of the two species separate them, with the  $ | e_{+} >
< e_{+} |$-specie occupying the left-half of the vessel and the $
| e_{-} > < e_{-} |$-specie occupying the right-half of the
vessel.
  \item finally an isothermal compression takes the system in a situation in which the volume and the pressure are the same of the initial state; such
  a compression requires an expenditure of work of:
\begin{equation}
  \Delta L _{2} \; = \; -  n T [ \rho_{1} \log  \rho_{1} \, +  \, \rho_{2} \log  \rho_{2} ]
\end{equation}
\item finally a suitable unitary evolution takes the system again  in the initial state.
\end{itemize}
The net work made by the engine during the cycle is:
\begin{equation}
   \Delta L \; = \; \Delta L _{1} + \Delta L _{2} \; > \; 0
\end{equation}
so that the whole thermodynamical cycle converts the heat
extracted by the reservoir in a positive amount of work of $
\Delta L $.

This, according to Peres, violates the Second Principle, proving
that \textbf{the "magic"  membranes able to separate nonorthogonal
states with perfect efficiency} cannot exist.

Such a proof, anyway, in not correct, owing to Zurek's theorem;
the key point touches the conceptual deepness underlying
eq.\ref{eq:Zurek's theorem}, whose complete comprehension
requires to explicitly analyze the bug in Von Neumann's proof
that $ S_{therm} ( \rho ) \, = \, I_{prob} ( \rho ) $.

The key point is based in the own definition of the
semi-permeables membranes of Einstein's method: as correctly
observed by Peres \textbf{the semipermeable-membranes are endowed
with automatic devices able to peak inside the boxes and to test
the state}.

What Peres seems unfortunately not to catch is that  a
semi-permeable membrane is  then an  \textbf{intelligent system}
operating in the following way:
\begin{enumerate}
  \item gets the input $ ( s , i ) $  from a device measuring both the side s from which the $ | \phi_{i} > < \phi_{i} | $-specie arrives and its kind, i.e. the classical information codified by its label i.
  \item computes a certain semaphore-function p such that $ ( s , i ) \stackrel{p}{\rightarrow}  p[(s,i)] $ giving as output a 0 if the $ | \phi_{i} > < \phi_{i} | $-specie must be
  left to pass while gives as output a one if the $ | \phi_{i} > < \phi_{i} | $-specie must be stopped
  \item gives the output p[(s,i)] to a suitable device that operates on the  $ | \phi_{i} > < \phi_{i} | $-specie in the specified way
\end{enumerate}
The argument of Bennett's exorcism concerning the necessity of
taking into account the  algorithmic-information of the sequences
of successive recorded $ ( s , i ) $'s in the membrane's memory
thus apply.

But this must be done, in particular, in the cases of
Peres'-engine:

taking into account also the algorithmic-information of the
semi-permeable's membranes, one sees that it is greater than or
equal to the universe's entropy decrease corresponding to the
work made by the engine, so that, by eq.\ref{eq:Zurek's theorem}:
\begin{equation}
  \Delta S_{therm} \; \geq \; 0
\end{equation}
and Peres' arguments falls down.

\end{document}